\documentclass{aastex}
\usepackage{emulateapj5,apjfonts}

\newcommand{\ltsim}{\raisebox{-.5ex}{$\;\stackrel{<}{\sim}\;$}}
\newcommand{\gtsim}{\raisebox{-.5ex}{$\;\stackrel{>}{\sim}\;$}}
\newcommand{\kms}{\ifmmode {\rm km\ s}^{-1} \else km s$^{-1}$\fi}
\newcommand{\et}{et al.\ }

\newcommand{\nvciv}{\ion{N}{5}/\ion{C}{4}}
\newcommand{\nvheii}{\ion{N}{5}/\ion{He}{2}}
\newcommand{\nivcivA}{\ion{N}{4}]~$\lambda1486$/\ion{C}{4}~$\lambda1549$}
\newcommand{\nivciv}{\ion{N}{4}]/\ion{C}{4}}
\newcommand{\oiiiciv}{\ion{O}{3}]~$\lambda1663$/\ion{C}{4}~$\lambda1549$}
\newcommand{\niiiciii}{\ion{N}{3}]~$\lambda1750$/\ion{C}{3}]~$\lambda1909$}
\newcommand{\niiioiii}{\ion{N}{3}]~$\lambda1750$/\ion{O}{3}]~$\lambda1663$}

\shortauthors{SHEMMER \& NETZER}
\shorttitle{METALLICITY--LUMINOSITY RELATIONSHIP IN AGNs}

\journalinfo{The Astrophysical Journal, 567:L19--L22, 2002 March 1,
astro-ph/0201437}

\slugcomment{Received 2001 November 14; accepted 2002 January 22; published 2002
February 8}

\begin{document}

\title{IS THERE A METALLICITY-LUMINOSITY RELATIONSHIP IN ACTIVE GALACTIC NUCLEI? \\
THE CASE OF NARROW-LINE SEYFERT 1 GALAXIES}

\author{
OHAD SHEMMER\altaffilmark{1}
AND HAGAI NETZER\altaffilmark{1}
}

\altaffiltext{1}
                {School of Physics and Astronomy and the Wise
                Observatory, Raymond and Beverly Sackler Faculty of
                Exact Sciences, Tel-Aviv University, Tel-Aviv 69978,
                Israel; ohad@wise.tau.ac.il, netzer@wise.tau.ac.il.}

\begin{abstract}

The well-known relationship between metallicity and luminosity in
active galactic nuclei (AGNs)
is addressed by introducing new metallicity measurements (based on the
method of Hamann \& Ferland, hereafter HF) for a sample of narrow-line
Seyfert 1 galaxies (NLS1s). Our new results, based on a sample of 162
AGNs, including nine NLS1s, indicate that while broad-line AGNs trace a
metallicity--luminosity power law with an index of $\sim0.2$, NLS1s
deviate significantly from this relationship at low luminosities. Adopting
the HF method based on the \nvciv\ line ratio, we find that NLS1
metallicities are similar to those of some high-redshift, high-luminosity quasars.
We also examined the \nivciv\ line ratio and compared it with  \nvciv\ in a sample
of 30 sources including several NLS1s. We find that the two do not give a
consistent answer regarding the N/C abundance ratio. This result is marginal
because of the quality of the data. We suggest two alternative explanations
to these results: 1) The HF metallicity--luminosity dependence is not a
simple two-parameter dependence and there is an additional hidden variable
in this relationship that has not yet been discovered. The additional
parameter may be the accretion rate, the age of the central stellar cluster
or, perhaps, something else. 2) The strong line ratios involving
\ion{N}{5}~$\lambda1240$ suggested by HF are not adequate metallicity
indicators for NLS1s and perhaps also other AGNs for reasons that are not yet
fully understood.

\end{abstract}


\section{THE METALLICITY-LUMINOSITY RELATIONSHIP IN ACTIVE GALACTIC NUCLEI
\label{ZLinAGN}}

Studies of emission lines in active galactic nuclei (AGNs) indicate
that metallicities are typically near the solar value in their broad-
line region (BLR). Accurate determinations are difficult to obtain
since the line ratios depend on unknown densities and optical depths
\cite{net90}. However, Shields (1976) showed that some BLR abundances
can be derived from the observed line ratios of weak nitrogen, carbon, and
oxygen lines,  almost independent of the physical properties of the gas
(see Hamann \et 2002 for more details). One such line ratio is \nivcivA\
with a theoretical value of $\sim0.045$ for solar metallicity.  Others are
\niiioiii, \oiiiciv\, and \niiiciii, which can determine the N/O, O/C,
and N/C abundances, respectively.  This method has been used in several
AGN studies (e.g., Baldwin \& Netzer 1978; Osmer 1980; Uomoto 1984),
but there are practical limitations due to the weakness of these lines.

An important development in this area was achieved by Hamann \& Ferland
(1993, hereafter HF93; see also Hamann \& Ferland 1999), who suggested
alternative abundance indicators that are somewhat model dependent
but much easier to obtain observationally. In particular, they have
shown that the \ion{N}{5}~$\lambda1240$/\ion{C}{4}~$\lambda1549$~and
\ion{N}{5}~$\lambda1240$/\ion{He}{2}~$\lambda1640$~line ratios represent
the overall BLR metallicity. They also claimed that BLR metallicity tends
to grow with AGN luminosity up to $\sim10$Z$_{\odot}$, thus implying a
metallicity--luminosity (hereafter $Z-L$) relationship, in analogy with
the mass--metallicity relationship observed for elliptical galaxies.
At one extreme end of this relationship one finds the luminous and high-
redshift quasars as AGNs having the highest metallicities. In several
cases, those quasars display relatively narrow UV emission lines
(e.g., Osmer 1980; Warner \et 2002).

The low-luminosity regime in the HF93 diagram is also occupied by narrow-line
Seyfert 1 galaxies (NLS1s) that have been mostly left out from previous abundance
analyses. These NLS1s are defined by their extremely
narrow optical permitted emission lines (FWHM$\ltsim2000$\,\kms) in comparison
with normal broad-line AGNs
(BLAGNs; Osterbrock \& Pogge 1985). NLS1s show extreme AGN properties;
their optical emission lines put them at one extreme end of the Boroson
\& Green (1992) primary eigenvector, and they tend to
display unusual behavior in other wave bands, especially in the X-ray
(e.g., Boller, Brandt, \& Fink 1996; Leighly 1999a, 1999b).  A possible
explanation for the peculiar properties of NLS1s is that they have
relatively low black hole (BH) masses for their luminosities and hence a
very large $L/L_{\rm Edd}$. Since only a handful of NLS1s had their BH mass
measured directly using reverberation mapping techniques (Peterson \et
2000), this is not yet fully confirmed. Another extreme NLS1 property
was recently suggested, namely, unusually high metallicities (Mathur 2000
and references therein).  However, to date, this evidence remains scarce,
and no systematic abundance study has yet been carried out.

In this study we present for the first time metallicity (\'{a} la HF93)
measurements for a sample of NLS1s. In \S~\ref{sample_prop} we define the
sample properties and data analysis. In \S~\ref{results} we describe
our new results on the $Z-L$ relationship in AGNs and attempt to answer two
related questions: (1) Do NLS1s have higher metallicities compared with
BLAGNs for a given luminosity? (2) Are the higher metallicities, assumed
for NLS1s related to the fundamental physical properties that drive the
NLS1 phenomenon?

\section{SAMPLE PROPERTIES AND DATA ANALYSIS \label{sample_prop}}

We selected 162 type 1 AGNs, including nine extreme NLS1s, that had
either published optical or UV ({\sl HST}) HF93 line ratios or an UV
({\sl HST}-archived) good-quality spectrum
\centerline{\includegraphics[width=8.5cm]{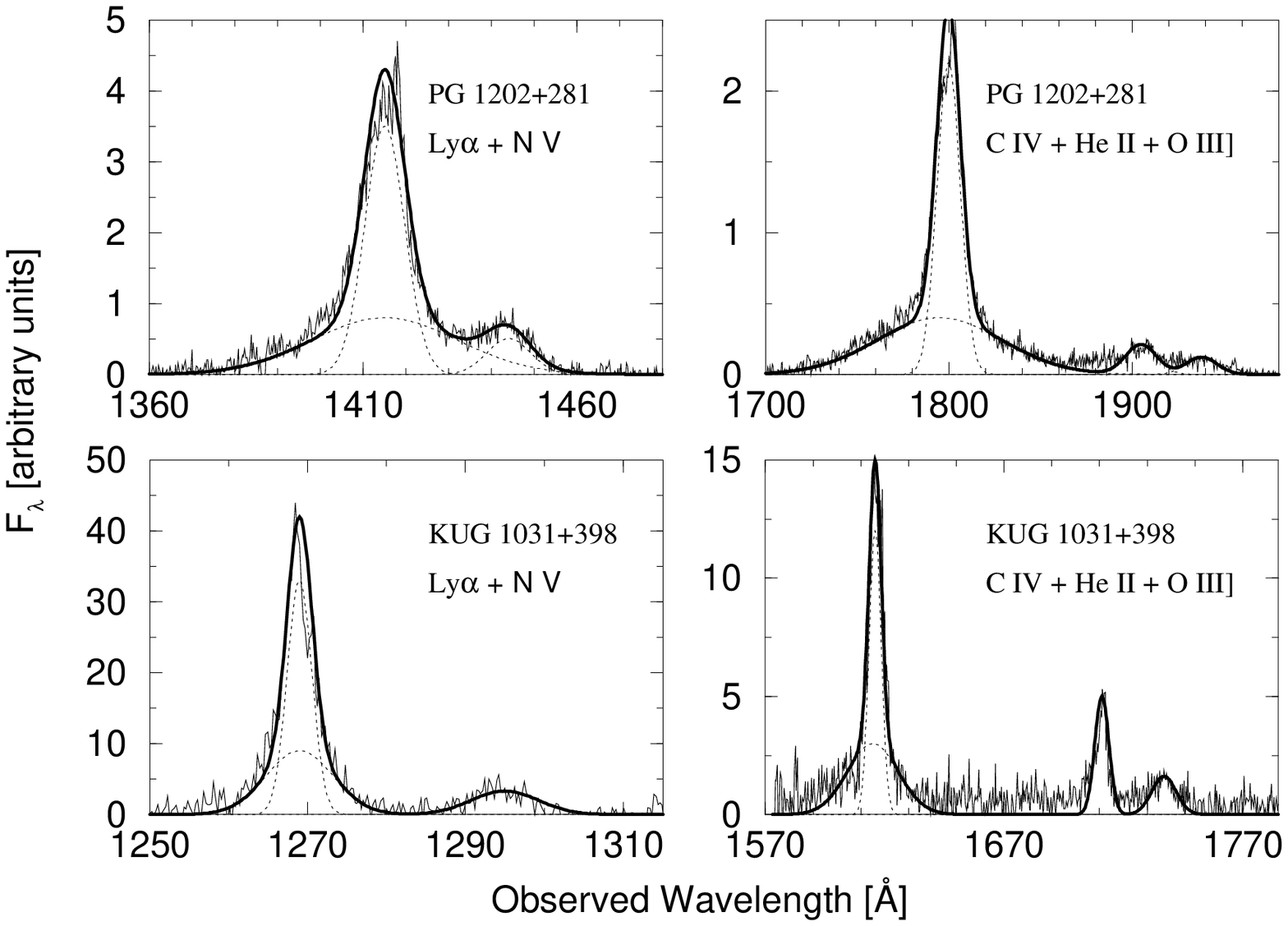}}
\figcaption{Two typical line-fitting examples: A BLAGN PG~1202$+$281 ({\it top panels})
and an NLS1 KUG 1031+398 ({\it bottom panels}). In each continuum-subtracted spectrum,
the Ly$\alpha$ region ({\it left panels}) and the C\,{\sc iv} region ({\it right panels})
were fitted as explained in the text. In each panel a thin solid line represents
the spectrum; dotted lines represent the individual Gaussian profiles that
are then summed to form the entire fit to the spectrum ({\it thick solid lines}).}
\label{fitting}
\centerline{}
\noindent covering the rest-frame
range of $\sim\lambda\lambda1200$--$1700$. We also inspected UV({\sl IUE})
spectra of 11 extreme NLS1s, but the poor signal-to-noise
ratios resulted in line flux uncertainties that were too large to be included
in our study (although the general trend was similar to the nine NLS1
spectra taken from {\sl HST}). A redshift of~z\,$\approx$\,1.6 defines the
transition from low-z objects, for which the relevant emission lines are
detected in the UV, to those objects for which \ion{N}{5} is detected
in the optical band and are hereafter referred to as high-z quasars. We
consider NLS1s as AGNs having FWHM(H$\beta$)\ltsim\,1500\,\kms in order to
concentrate on the more extreme objects of this class \footnote{H$\beta$
is the most suitable emission line for distinguishing broad-lined BLRs
from narrow-lined ones. Measurements of FWHM(H$\beta$) are not available for
$\sim60$\% of our sample's objects that have z\gtsim0.5.
Throughout this Letter, we treat those unknown FWHM(H$\beta$)
objects as BLAGNs.}.

Among the 162 objects, 67 have published optical spectral data and
are regarded as high-z and unknown FWHM(H$\beta$) objects (Osmer
\& Smith 1976, 1977; Baldwin \& Netzer 1978; Osmer 1980; Uomoto 1984; Baldwin,
Wampler, \& Gaskell 1989; Baldwin \et 1996; Dietrich \&
Wilhelm-Erkens 2000). All those include detailed line intensity measurements.
The rest, 95 low-z objects, including the nine extreme NLS1s, have UV ({\sl HST})
spectra. Of these, 44 have published line intensities
(Reichert \et 1994; Laor \et 1994, 1995, 1997; Wills \et 1995).
Emission-line fluxes of 111 objects were thus readily available for our analysis.
For the remaining 51 objects, we obtained the HF93 line ratios
by performing line and continuum measurements on the archived spectra
by applying a multi-Gaussian component fit using the {\sc{ngaussfit}}
task in IRAF\footnote{{IRAF (Image Reduction and Analysis Facility) is
distributed by the National Optical Astronomy Observatories, which are
operated by AURA, Inc., under cooperative agreement with the National
Science Foundation.}}. For Ly$\alpha$ and \ion{C}{4}~$\lambda1549$,
two Gaussian components were fitted, while for \ion{N}{5}~$\lambda1240$,
\ion{He}{2}~$\lambda1640$, and \ion{O}{3}]~$\lambda1663$ we fitted a
single Gaussian. Two typical fits of this kind appear in Figure~1.
The individual fits produced uncertainties on the line
fluxes that were much smaller compared with the uncertainties introduced
by other unknowns, such as the precise continuum placement and slope,
line contamination, etc. In particular, the \ion{Si}{2}~$\lambda1264$ complex
might be expected to contribute at most $\sim20$\% (Laor \et 1997) to the
\ion{N}{5}~$\lambda1240$ intensity only in AGNs with FWHM(\ion{N}{5})$\gtsim
5000$~\kms. In light of the underestimated uncertainties,
we compared published flux values with new measurements that we performed for
several
\centerline{}
\centerline{\includegraphics[width=8.5cm]{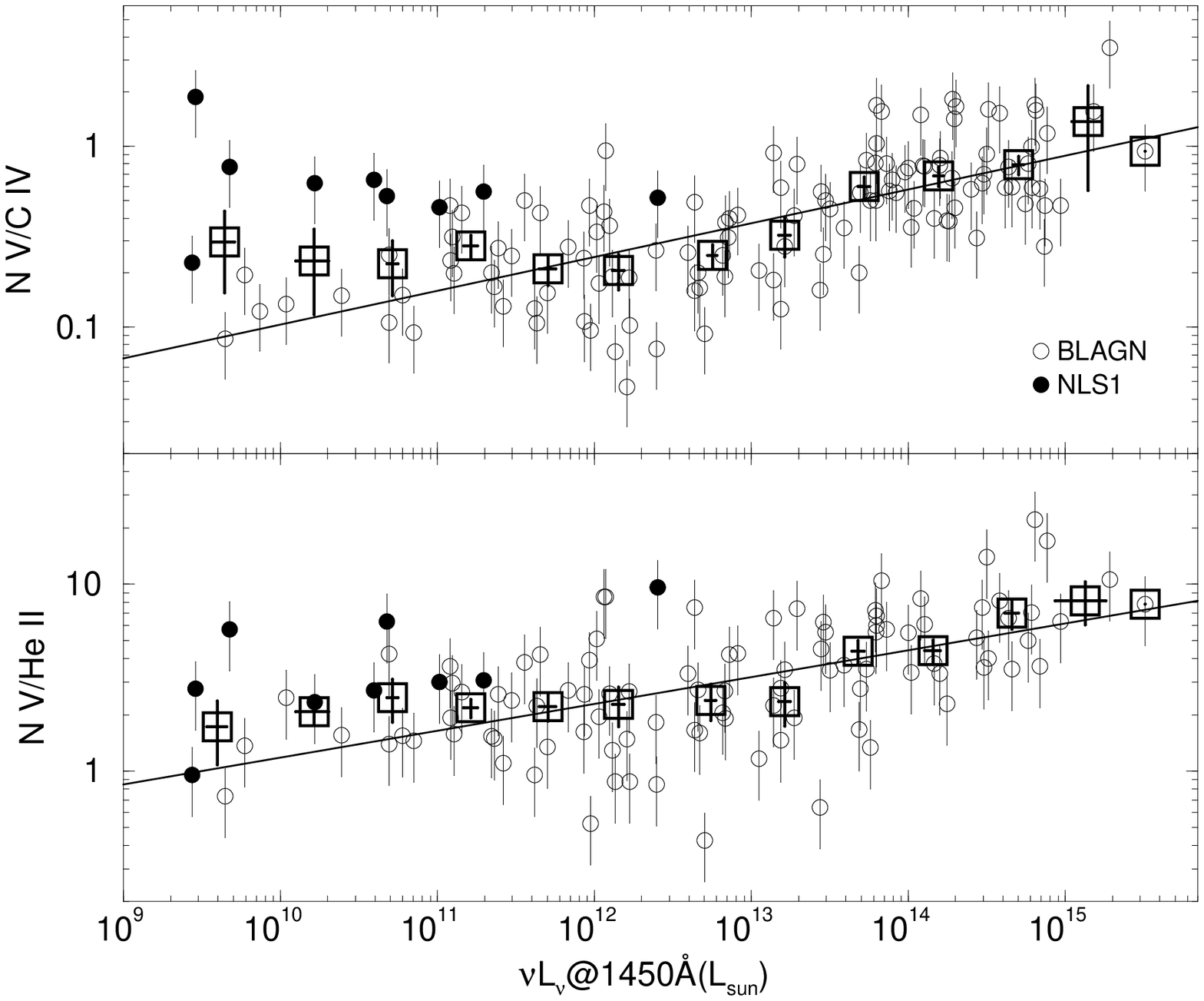}}
\figcaption{HF93 metallicity indicators, N\,{\sc v}/C\,{\sc iv}
({\it top}) and N\,{\sc v}/He\,{\sc ii} ({\it bottom}), as a function
of luminosity. Empty circles mark BLAGNs, and filled circles marks NLS1s;
solid lines represent the BLAGN best-fit $Z-L$ slope; large squares
with error bars represent average line ratios in bins of 0.5 in log
$\nu L_{\nu}$ of the entire data set. Note the significant deviation of
the low-luminosity bins from the straight line ({\it top}) owing to the
addition of NLS1s.}
\label{ratios}
\centerline{}
\noindent test cases and concluded that uncertainties on the line fluxes we
measured can nominally be taken as 30\%. Finally, we included only
positive measurements of line intensities, excluding a few cases
of upper flux limits, that did not alter our basic finding as discussed below.
We were therefore left with 130 (107) positive measures of the \nvciv\, (\nvheii)
line ratios. Luminosities were derived from the monochromatic
flux at $\lambda1450$, taking $H_0=75$ \kms~Mpc$^{-1}$ and $q_0=0.5$.

Our sample is far from being statistically complete. However,
we do not know of any observational or selection effect that will bias the
sample against strong \ion{N}{5} lines (that are easily detectable and measurable).
In particular, our sample of UV ({\sl HST}) sources is strongly biased
in favor of radio-loud quasars (e.g., Wills \et 1995). To test this effect,
we carried out the analyses as discussed below while excluding line
ratios of the radio-loud population. Our results were not significantly
affected by this test (see \S~\ref{results}).
Thus, while the following correlation slopes may depend on the specific
sample, the important result (the fact that low-luminosity sources
show very strong \nvciv; see \S~\ref{results}) is insensitive
to that.

\section{RESULTS AND DISCUSSION \label{results}}
\subsection{New Correlations Involving \ion{N}{5}~$\lambda1240$ \label{N5}}

Figure~\ref{ratios} shows the \nvciv\ ({\it top}) and \nvheii\
({\it bottom}) line ratios as a function of luminosity.  The HF93
line ratios of the entire sample were also binned in ranges of 0.5
in log $\nu L_{\nu}$ to minimize the effect of uneven distribution
in luminosity. The average positions of all objects in each of those
bins are shown as large squares with error bars in Figure~\ref{ratios}.
The error bars on the large squares represent, for each axis, the standard
deviation divided by the square root of the number of objects in each bin.
For each HF93 line ratio--$\nu L_{\nu}$ log-log diagram, we performed a
linear regression analysis and calculated the Pearson and
Spear-

\scriptsize
\begin{center}
{\sc TABLE 1 \\ Linear Regression Parameters for N\,{\sc v}/C\,{\sc iv} versus log $\nu L_{\nu}$}
\vskip 4pt
\begin{tabular}{lccccc}
\hline
\hline
{Data Set} &
{Number of} &
{Pearson} &
{Spearman} &
{Slope} &
{Constant} \\
{Code$\rm ^a$} &
{Objects} &
{($r$)} &
{($r_s$)} &
{($a$)} &
{($b$)} \\
\hline
B           & 121 & 0.70 & 0.73 & $0.19\pm0.02$ & $-2.86\pm0.23$  \\
B$+$N       & 130 & 0.55 & 0.60 & $0.13\pm0.02$ & $-2.11\pm0.23$  \\
RQQ (B)     & 105 & 0.72 & 0.74 & $0.18\pm0.02$ & $-2.82\pm0.23$  \\
RQQ (B$+$N) & 114 & 0.56 & 0.60 & $0.13\pm0.02$ & $-2.04\pm0.24$  \\
B$+$up.lim. & 137 & 0.67 & 0.69 & $0.19\pm0.02$ & $-2.91\pm0.24$  \\
B$+$N$+$up.lim. & 146 & 0.52 & 0.57 & $0.13\pm0.02$ & $-2.17\pm0.24$  \\
\hline
\end{tabular}
\vskip 2pt
\parbox{3.485in}{
\small\baselineskip 9pt
\footnotesize
\indent
$\rm ^a${Data set codes are B for BLAGNs, N for NLS1s,
RQQ for radio-quiet quasars, and up.lim. for upper limits on the line ratio.}
}
\end{center}
\setcounter{table}{1}
\normalsize

\scriptsize
\begin{center}
{\sc TABLE 2 \\ Linear Regression Parameters for N\,{\sc v}/He\,{\sc ii} versus log $\nu L_{\nu}$}
\vskip 4pt
\begin{tabular}{lccccc}
\hline
\hline
{Data Set} &
{Number of} &
{Pearson} &
{Spearman} &
{Slope} &
{Constant} \\
{Code$\rm ^a$} &
{Objects} &
{($r$)} &
{($r_s$)} &
{($a$)} &
{($b$)} \\
\hline
B          &  98 & 0.58 & 0.60 & $0.14\pm0.02$ & $-1.37\pm0.27$  \\
B$+$N      & 107 & 0.50 & 0.53 & $0.11\pm0.02$ & $-0.95\pm0.24$  \\
RQQ (B)    &  83 & 0.61 & 0.64 & $0.14\pm0.02$ & $-1.25\pm0.26$  \\
RQQ (B$+$N)&  92 & 0.54 & 0.57 & $0.11\pm0.02$ & $-0.87\pm0.23$  \\
B$+$up.lim.& 110 & 0.54 & 0.55 & $0.15\pm0.02$ & $-1.45\pm0.28$  \\
B$+$N$+$up.lim.& 119 & 0.46 & 0.48 & $0.11\pm0.02$ & $-1.00\pm0.26$  \\
\hline
\end{tabular}
\vskip 2pt
\parbox{3.485in}{
\small\baselineskip 9pt
\footnotesize
\indent
$\rm ^a${Data set codes are identical to those in Table 1.}
}
\end{center}
\setcounter{table}{2}
\normalsize

\noindent man linear correlation coefficients. This analysis was carried out once for
the BLAGN population and a second time for the entire data set. The results are
presented in Tables 1 and 2; one can see that the exclusion of radio-loud objects
or the addition of flux ratio upper limits (that were treated as real ratios) had
little effect on the correlations and slopes.

Inspection of Figure~\ref{ratios} and Tables 1 and 2
shows that the \nvciv\, ratio is strongly correlated with luminosity in the
BLAGN case. However, it is also apparent that this relationship {\it breaks
down completely} at low luminosities, when the NLS1s are introduced into
the sample. Several NLS1s that belong in the lowest luminosity regime
even have \nvciv\, ratios as high as those of the most luminous high-z
quasars. The extremely strong \ion{N}{5}~$\lambda1240$ in some NLS1s has
been noted earlier by Wills et al. (1999) who did not investigate the
resulting N/C abundance ratio. The \nvheii\, ratio behaves somewhat
differently when NLS1s are added (Fig.~\ref{ratios}). We find that the
\nvheii\, ratio in BLAGNs is correlated with luminosity, although not as
strong as in the \nvciv\, case, and that NLS1s only slightly deviate from
the \nvheii\, $Z-L$ slope. We also discover that both line ratios are not
correlated with luminosity for $\nu L_{\nu}$ at $1450$\AA\,$\ltsim10^{46}$ erg s$^{-1}$,
neither for BLAGNs nor for the entire sample. We emphasize once more that
the \nvciv\ ratio in NLS1s is very similar to the one observed in high-$L$
AGNs and this result, which is based on an easy-to-measure line ratio,
is enough to completely change the original HF93 correlation, regardless
of the exact slope or value of the correlation coefficient.

\subsection{High \nvciv\ at Low Luminosity \label{reliable}}

Our new results point at two distinct scenarios: either (1) the HF93 line
ratios overpredict N/C at least under some physical conditions or (2)
high metallicities at low luminosities are possible and are seen in NLS1s.
We discuss briefly the implications of these two scenarios and defer the
more detailed analysis to a later publication.

\subsubsection{Is the \nvciv\ Line Ratio a Reliable N/C Indicator?}

Hamann \et (2002, hereafter H02) have used state-of-the-art photoionization
calculations to investigate several line ratios in order to select those that
are robust abundance indicators. The calculations span a vast range of densities
[$7\leq$ log $n_{\rm H} (cm^{-3})\leq 14$] that completely cover the typical range
attributed to the broad emission line gas in AGNs [log $n_{\rm H} (cm^{-3}) \approx10$].
Since there is no evidence for densities as large as $10^{13}$--$10^{14} \ cm^{-3}$
in NLS1s, we have no reason to suspect that they lie outside the range covered by
the H02 calculations. According to the calculations, the \nvciv\ line ratio is a reliable
N/C indicator over the range of interesting physical conditions expected in the BLR.
H02 have also considered several of the weak intercombination lines, such as \niiioiii\
and \nivciv\ , previously discussed by Shields (1976), and concluded that the first is
more reliable than the second, as it is less sensitive to the model assumptions. Results
for all relevant line ratios are shown in their Figure~4 and a specific application
to the locally optimally emitting clouds (LOCs) model is shown in H02 Figure~5.

The only other line ratio available to us, except for the \nvciv\ ratio discussed above,
is \nivciv. We have therefore investigated the N/C obtained from the two line pairs
both observationally and theoretically. Several BLAGNs of our sample had published flux
values (or upper flux limits) of the weak line \ion{N}{4}]. We added to
this subsample measurements of the \ion{N}{4}] line in eight of our nine
extreme NLS1s.  The N/C abundance obtained from the \nivciv\ line ratio
was then calculated for the subsample, assuming that upper limits represent real
ratios. We find that the two line ratios are well correlated ($r=0.8$ for 30 sources);
i.e., if \nvciv\ is a good N/C indicator, so is \nivciv. However, the derived N/C,
{\it assuming both ratios are reliable metallicity indicators}, is very different.
The \nvciv\ ratio gives systematically larger N/C, sometimes by a factor as large as
3 or 4. We stress that the results are still tentative because of the large number
of upper limits rather than real line ratios used in the analysis. Better data are
required to confirm this correlation.

Regarding the suitability of \nivciv\ as an N/C indicator, we note that the
calculations presented in Figure 4 of H02 show that the line ratio does not
change by more than a factor of 2 over the range of conditions thought to
be acceptable in AGN BLRs (ionization parameter of 0.03--0.3 for the H02 continuum and
density below about 10$^{12}$ cm$^{-3}$). The H02 conclusion that the line ratio is not
a robust N/C indicator is based on regions in parameter space that are different from
the one specified above. Moreover, the particular example shown in Figure 5 of H02, applicable to
the LOC model, clearly shows that under such conditions (that produce well most of the
observed line ratios in AGNs; see Baldwin et al. 1995), the \nivciv\ line ratio is
indeed a very good N/C indicator. A key issue is whether or not the physical conditions
in the BLR of NLS1s are similar to those in broader line AGNs. The idea that density and
optical depths may be different has been proposed in the past, and a better assessment of
the \nivciv\ suitability in this case must await a more detailed theoretical investigation
of such sources. At present we do not have a large enough sample and good enough
calculations to test the suggestion that the results shown in Figure~2 are due to
\nvciv\ being an inadequate N/C indicator.

\subsubsection{High-Metallicity NLS1s? \label{NLS1Z}}

The second scenario is based on the assumption that the \nvciv\ is a
reliable N/C indicator. This led HF93 to suggest a strong $Z-L$ relationship
in AGNs. However, our new measurements clearly show that NLS1s do not follow
this $Z-L$ relationship. NLS1 metallicities, as indicated by the \nvciv\ ratio,
are similar to those of the most luminous high-z quasars in our sample and
are higher, by almost an order of magnitude, than those of BLAGNs with similar
luminosities. In the \nvheii\ case, NLS1s show only slightly higher metallicities
for a given luminosity compared with BLAGNs. This effect may be attributed
to the more complex dependence of \nvheii\ on other physical parameters,
such as the spectral energy distribution. Accepting this scenario, the HF93
$Z-L$ relationship cannot be a simple two-parameter dependence for all AGNs.

We also attempted to find a link between BLAGNs and NLS1s in order to see
whether the $Z-L$ relationship is a smooth function of FWHM(H$\beta$).
We found no correlation between FWHM(H$\beta$) and metallicity,
luminosity, or a combination of the two. Since we have FWHM(H$\beta$)
values for about half of our sample, we checked whether FWHM(\ion{C}{4})
could be more suitable, since this emission line is more dominant in
our case. Again we found no correlation with the other parameters, nor any
FWHM(\ion{C}{4})--FWHM(H$\beta$) correlation, for objects
having both lines measured. In fact, we find that only AGNs that have
FWHM(H$\beta$)\ltsim\,1500 \kms\, follow a significantly different
$Z-L$ relationship.

If high metallicity is indeed another NLS1 extreme property, the
question whether this is related to some fundamental NLS1 physical
parameter remains unanswered. The introduction of NLS1s to the \nvciv\
$Z-L$ diagram completely changes the correlation at low $L$ and implies
that there is an additional dimension to this dependence that allows
high metallicities at low luminosities. This hidden variable in the
$Z-L$ relationship may be related to fundamental physical properties,
such as the accretion rate or the age of the central BH. The key to
answering this question possibly lies in the claim that NLS1s have low
BH masses for their luminosities (equivalent to larger $L/L_{\rm Edd}$ or
higher accretion rate) and therefore follow a different mass-luminosity
relationship than BLAGNs \cite{pet00}. Boroson \& Green (1992)
pointed out that $L/L_{\rm Edd}$
might be the underlying fundamental physical property of their primary
eigenvector. Since NLS1s lie at one extreme end of this eigenvector,
a natural hypothesis is that this property is also driving their
unusually high metallicities. In order to test this hypothesis, one
needs accurate BH mass determinations for our sample. Unfortunately,
those are available for only 34 objects \cite{kas00}, including only
three extreme NLS1s. Moreover, measuring BH masses for high-z quasars
is not a straightforward task since reliable determinations rely on
reverberation mapping studies that would require at least a decade-long
monitoring campaign due to cosmological time dilation. In
addition, measurements of [\ion{O}{3}], optical \ion{Fe}{2},
and H$\beta$ lines in high-z quasars are scarce (e.g. McIntosh \et 1999).

\subsection{The $Z-L$ Relationship in BLAGNs \label{ZL}}

Finally, we remain within the framework of the second scenario, where
the HF93 line ratios are considered reliable abundance indicators
and NLS1s show systematically higher metallicities than BLAGNs with
comparable luminosities.  We therefore removed all NLS1s from the sample
and checked the $Z-L$ relationship for BLAGNs [we caution the reader that
the BLAGN group includes objects with unknown FWHM(H$\beta$); see
\S~\ref{sample_prop}]. Our results confirm the observational evidence
for a correlation between metallicity and luminosity (HF93) for a large
(statistically incomplete) sample of BLAGNs. This correlation may be
written in the form $Z \propto L^{\alpha}$, where $Z$ and $L$ are the BLR
metallicity and BLAGN luminosity, respectively. Since both \nvciv\, and
\nvheii\, are approximately proportional to $Z$ (HF93), the index is
assigned with the mean value obtained for our sample, which is $\alpha \sim0.2$
(Tables 1 and 2). Combining this result with the empirical AGN
BH mass-luminosity relationship \cite{kas00}, $M_{\rm BH} \propto L^{0.5}$,
we find $Z \propto {M_{\rm BH}}^{0.4}$. Alternatively, one may argue that
the \cite{kas00} result regarding the $L$ versus $M$ relation is biased because
of the small size of the sample, and a more realistic form is perhaps
$M_{\rm BH} \propto L$. In this case, $Z \propto {M_{\rm BH}}^{0.2}$.
These results differ from
those that were reached by supermassive BH growth considerations where
$\alpha$ was assumed to take a value of $\approx\,0.5$ \cite{wan01}.
We note, again, that the above obtained $\alpha$ depends on the sample
selection and is, therefore, rather uncertain.

\acknowledgments

This research is supported in part by a grant from the Israel Science
Foundation. We thank an anonymous referee for useful comments and suggestions.

\end{document}